\RequirePackage{fix-cm}
\documentclass[twocolumn]{svjour3}          
\smartqed  
\usepackage{graphicx}
\usepackage{listings}
\usepackage{amsmath}
\usepackage{textcomp}

\usepackage[utf8]{inputenc}
\usepackage{tabularx}
\usepackage{xspace}
\usepackage{pifont}
\usepackage{epstopdf}
\usepackage{amssymb}
\usepackage{gensymb}
\usepackage{url}
\usepackage{multirow}
\usepackage{ctable}
\graphicspath{{./plots/}}
\usepackage{adjustbox,lipsum}
\usepackage{longtable}

\begin{document}

\title{Enhancing Regression Models for Complex Systems using Evolutionary Techniques for Feature Engineering}




\author{Patricia~Arroba         \and
        José~L.~Risco-Martín	\and
	Marina~Zapater		\and
	José~M.~Moya		\and
	José~L.~Ayala		
}


\institute{P. Arroba, J. M. Moya \at
              Electronic Engineering Dept.,\\
	      Technical University of Madrid, Madrid 28040, Spain\\
              \email{\{parroba,josem\}@die.upm.es}           
           \and
           J. L. Risco-Martín, J. L. Ayala \at
              DACYA, Complutense University of Madrid,\\
	      Madrid 28040, Spain\\
	      \email{\{jlrisco,jayala\}@ucm.es}
	   \and
	   M. Zapater \at
	      CEI Campus Moncloa\\
              UCM-UPM, Madrid 28040, Spain\\
              \email{marina@die.upm.es}
}

\date{Received: date / Accepted: date}

\maketitle

\begin{abstract}
This work proposes an automatic methodology for modeling complex systems.
Our methodology is based on the combination of Grammatical
Evolution and classical regression to obtain an optimal set of features that
take part of a linear and convex model.  This technique provides both Feature
Engineering and Symbolic Regression in order to infer accurate models with no
effort or designer's expertise requirements.  As advanced Cloud services are
becoming mainstream, the contribution of data centers in the overall power
consumption of modern cities is growing dramatically. These facilities consume
from 10 to 100 times more power per square foot than typical office buildings.
Modeling the power consumption for these infrastructures is crucial to
anticipate the effects of aggressive optimization policies, but accurate and
fast power modeling is a complex challenge for high-end servers not yet
satisfied by analytical approaches.  For this case study, our methodology
minimizes error in power prediction.  This work has been tested using real
Cloud applications resulting on an average error in power estimation of 3.98\%.
Our work improves the possibilities of deriving Cloud energy efficient policies
in Cloud data centers being applicable to other computing environments with
similar characteristics.

\end{abstract}

\section{Introduction}
\label{intro}

Analytical models, as closed form solution representations, require specific
knowledge about the different contributions and their relationships, becoming
hard and time-consuming techniques for describing complex systems.  Complex
systems comprise a high number of interacting variables, so the association
between their components is hard to extract and understand as they have
non-linearity characteristics~\cite{bar1997dynamics}.  Also, input parameter
limitations are barriers associated to classical modeling for these kind of
problems.  

Otherwise, classical regressions as least absolute shrinkage and
selection operator techniques provide models with linearity, convexity
and differentiability attributes, which are highly appreciated for describing
systems performance.  However, the automatic generation of accurate models for
complex systems is a difficult challenge that designers have not yet fulfilled
by using analytical approaches.

On the other hand, metaheuristics are higher-level procedures that
make few assumptions about the optimization problem, providing
adequately good solutions that could be based on fragmentary
information~\cite{Bianchi:2009:SMS:1541534.1541554,Blum:2003:MCO:937503.937505}.
They are particularly useful in solving optimization problems that are
noisy, irregular and change over time. In this way, metaheuristics
appear as a suitable approach to meet optimization problem
requirements for complex systems.

Some metaheuristics, as Genetic Programming (GP), perform Feature Engineering (FE) that
is a particularly useful technique for selecting an optimal set of features
that best describe an optimization problem.  Those features consist of
measurable properties or explanatory variables of a phenomenon.
FE methods select adequate characteristics avoiding the
inclusion of irrelevant parameters that reduce problem
generalization~\cite{ReidTurner19993}.  Finding relevant features
typically helps with prediction; but correlations and combinations of
representative variables, also provided by FE, may
offer a straightforward view of the problem thus generating better
solutions.

Grammatical Evolution (GE) is an evolutionary computation technique based on
GP. This technique is particularly useful to solve optimization problems and
provides solutions that include non-linear terms offering FE capabilities that
remove analytical modeling barriers.  One of the main characteristics of GE is
that it can be used to perform Symbolic Regression (SR)~\cite{GenProg}.  Also,
designer's expertise is not required to process a high volume of data as GE is
an automatic method. However, GE provides a vast space of solutions that may be
bounded to achieve algorithm efficiency.

In this work we propose a novel methodology for the automatic
inference of accurate models that combines the benefits offered by
both classic and evolutionary strategies.  Firstly, SR
performed by a GE algorithm finds
optimal sets of features that best describe the system behavior.
Then, a classic regression is used to solve our optimization problem
using this set of features providing the model coefficients. Finally,
our approach provides an accurate model that is linear, convex and
derivative and also uses the optimal set of features.  This
methodology can be applied to a broad set of optimization problems of
complex systems. This paper presents a case study for its application
in the area of Cloud power modeling as it is a relevant challenge
nowadays.

\subsection{Motivation}
One of the big challenges in data centers is the power-efficient
management of system resources. Data centers consume from 10 to 100
times more power per square foot than typical office
buildings~\cite{Scheihing:CreatingEnergyEfficient:07} even consuming
as much electricity as a city~\cite{Markoff:Intel:02}. Consequently, a
careful management of the power consumption in these infrastructures
is required to drive the Green Cloud
computing~\cite{Buyya:EnergyEfficientManagement:10}.

Cloud computing addresses the problem of costly computing infrastructures by
providing dynamic resource provision on a pay-as-you-go basis, and nowadays it
is considered as a valid alternative to owned high performance computing (HPC)
clusters. There are two main appealing incentives for this emerging paradigm:
firstly, the Clouds’ utility-based usage model allows clients to pay per use,
increasing the user satisfaction; secondly, there is only a relatively low
investment required for the remote devices that access the Cloud
resources~\cite{Chen:ProfilingVMs:11}.

Besides economic incentives, the Cloud model provides also benefits from the
environmental perspective, since the computing resources are managed by Cloud
service providers but shared among all users, which increases their overall
utilization~\cite{Berl:EECC:2010}. This fact is translated into a reduced
carbon footprint per executed task, diminishing $CO_2$ emissions. The Schneider
Electric’s report on virtualization and Cloud computing
efficiency~\cite{SchneiderReport} confirms that about 17\% of annual savings in
energy consumption were achieved by 2011 through virtualization technologies.

However, the proliferation of modern data centers is growing massively
due to the current increase of applications offered through the Cloud.
A single data center, that houses the computer systems and resources
needed to offer these services, has a power consumption comparable to
25000 households~\cite{Kaplan_Forrest_Kindler_2008}.  As a
consequence, the contribution of data centers in the overall
consumption of modern cities is increasing dramatically. Therefore,
minimizing the energy consumption of these infrastructures is a major
challenge to reduce both environmental and economic impact.

The management of energy-efficient techniques and aggressive
optimization policies requires a reliable prediction of the effects
caused by the different procedures throughout the data center.  Server
heterogeneity and diversity of data center configurations difficult to
infer general models.  Also, power dependency with non-traditional
factors (like the static consumption and its dependence on
temperature, among others) that affect consumption patterns of these
facilities may be devised in order to achieve accurate power models.

These power models facilitate the analysis of several architectures
from the perspective of the power consumption, and allow to devise
efficient techniques for energy optimization.  Data center designers
have collided with the lack of accurate power models for the
energy-efficient provisioning and the real-time management of the
computing facilities.  Therefore, a fast and accurate method is
required to achieve overall power consumption prediction.

The work proposed in this paper makes substantial contributions in the
area of power modeling of Cloud servers taking into account these
factors. We envision a powerful method for the automatic
identification of fast and accurate power models that target high-end
Cloud server architectures. Our methodology considers the main sources
of power consumption as well as the architecture-dependent parameters
that drive today's most relevant optimization policies.

\subsection{Contributions}
Our work makes the following contributions:

\begin{itemize}
\item We propose a method for the automatic generation of fast and
  accurate models adapted to the behavior of complex systems.

\item Resulting models include combination and correlation of
  variables due to the FE and SR
  performed by GE.  Therefore, the models
  incorporate the optimal selection of representative features that
  best describe system performance.

\item Through the combination of GE and classical
  regression provided by our approach, the inferred models have
  linearity, convexity and differentiability properties.

\item As a case study, different power models have been built and tested for
  a high-end server architecture running several real applications
  that can be commonly found in nowadays’ Cloud data centers,
  achieving low error when compared to real measurements.

\item Testing for different applications (web search engines, and
  both memory and CPU-intensive applications) shows an average error
  of 3.98\% in power estimation.
\end{itemize}

The remainder of this paper is organized as follows:
Section~\ref{relWork} gives further information on the related work on
this topic.  Section~\ref{algorithm} provides the background
algorithms used for the model optimization.  The methodology
description is presented in Section~\ref{probMod}.  In
Section~\ref{caseStudy} we provide a case study where our optimization
modeling methodology is applied.  Section~\ref{results} describes
profusely the experimental results.  Finally, in
Section~\ref{conclusions} the main conclusions are drawn.

\section{Related Work}
\label{relWork}

A complex system can be described as an interconnected agents system exhibiting
a global behavior that results from agents
interactions~\cite{boccara2010modeling}.  Nowadays, the number of agents in a
system grows in complexity, from data traffic scenarios to multisensor systems,
as well as the possible interactions between them. Therefore, infering the
global behavior, not imposed by a central controller, is a complex and
time-consuming challenge that requires a deep knowledge of the system
performance.
Due of these facts, new automatic techniques are required to facilitate the
fast generation of models that are suitable for complex systems presenting a
large number of variables.

The case study presented in this work exhibits high complexity in terms of
number of variables and possible traditional and non-traditional sources of
power consumption.  This issue demands the following review of the
state-of-the-art.

In the last years, there has been a rising interest in developing
simple techniques that provide basic power management for servers
operating in a Cloud, i.e.  turning on and off servers, putting them
to sleep or using Dynamic Voltage and Frequency Scaling (DVFS) to
adjust servers’ power states by reducing clock frequency. Many of
these recent research works have focused on reducing power consumption
in cluster systems
~\cite{Adiga:BlueGene:02},~\cite{Warren:HighDensityComputing:02},~\cite{Ge:2005:PDD},~\cite{Hsu:PowerAware:05}.

In general, these techniques take advantage of the fact that
application performance can be adjusted to utilize idle time on the
processor to save energy~\cite{Contreras:2005:PPI}. However, their
application in Cloud servers is difficult to achieve in practice as
the service provider usually over-provisions its power capacity to
address worst case scenarios. This often results in either waste of
power or severe under-utilization of resources.  Thus, it is critical
to quantitatively understand the relationship between power
consumption, temperature and load at the system level by the
development of a power model that helps in optimizing the use of the
deployed Cloud services.

Currently the state-of-the-art offers various analytical power
models. However, these models are architecture-dependant and do not
include the contribution of static power consumption, or the
capability of switching the frequency modes (DVFS). The authors
develop linear regression models that present the power consumption of
a server as a linear function of its CPU
usage~\cite{Lewis:2008:REC},~\cite{Pelley:UnderstandingAbstracting:09}. Some
other models can be found where server power is formulated as a
quadratic function of the CPU
usage~\cite{Meisner:2010:PPM},~\cite{Warkozek:NewApproachModel:12}. Still,
as opposed to ours, these models do not include the estimation of the
static power consumption (which has turned to have a great impact due
to the current server technology).

Bohra et al.~\cite{Bohra:ParallelDistributedProcessing:10} propose a
robust fitting to calculate their model that takes into account the
correlation between the total system power consumption and component
utilization. Our work follows a similar approach but also incorporates
the contribution of the static power consumption, its dependence on
temperature, and the effect of applying voltage and frequency scaling
techniques.

Interestingly, one key aspect in the management of a data center is
still not very well understood: controlling the ambient temperature at
which the data center operates. Data centers operate in a broad
temperature range from 18$^\circ$C to 24$^\circ$C but some can be as
cold as
13$^\circ$C~\cite{Brandon:GoingGreenData:07,Miller:Google:08}. However,
due to the lack of accurate power models, the effect of ambient
temperature on the power consumption of the servers has not been
clearly analyzed, preventing the application of optimization models to
save energy. On the contrary, the experimental work presented in this
paper has been performed in ambient temperatures ranging from
18$^\circ$C to 25$^\circ$C. The range selected follows nowadays'
practice of operating at higher temperatures ~\cite{google:tempDC} and
close to the limits recommended by ASHRAE. Although this practice
obtains energy savings in the cooling expense~\cite{El-Sayed:2012},
the lack of a detailed power model prevents to apply optimization
policies.
  
In our previous work, we have applied the benefits of Particle Swarm
Optimization algorithms (PSO) to identify an analytical model that
provides accurate results for power estimation~\cite{parrobaSEB}.  PSO
simplifies the power model by significantly reducing the number of
predefined parameters and variables used in the analytical
formulation. However, as a parameter identification mechanism, this
technique does not provide an optimal search of the features that best
describe the system power performance, so additional features could be
incorporated.

The work presented in this paper outperforms previous approaches in
the area of power modeling for enterprise servers in Cloud facilities
in several aspects:

\begin{itemize}
\item Our approach consists on an automatic method for the
  identification of an accurate power model particularized for each
  target architecture. We propose an extensive power model consistent
  with current architectures.

\item The proposed methodology takes into account the main power
  consumption sources resulting in a multiparametric model to allow
  the development of power optimization approaches. Different
  parameters are combined by Feature engineering assuring that the
  optimal set of features is considered.

\item Optimal features are included in a classical regression
  resulting in a specific model instance for every target architecture
  that is linear, convex and derivable. Also the execution of the
  resulting power model is fast, making it suitable for run-time
  optimization techniques.
\end{itemize}

\section{Algorithm description}
\label{algorithm}

\subsection{Grammatical Evolution}
As previously stated, we work on FE to obtain mathematical expressions
that represent different power consumption contributions.  These
expressions, or features, are derived from the combination of
previously collected experimental parameters (in our case of study,
they correspond to processor and memory temperatures, fan speeds,
processor and memory utilizations, processor frequencies
and voltages). We deal with a kind of SR problem
to select the relevant features. SR tries to simultaneously obtain a
mathematical expression while including the relevant parameters to
reproduce a set of discrete data. Besides, GP
has proven to be effective in a number of SR
problems~\cite{Vladislavleva2009}. However, GP presents some
limitations like bloating of the evolution (excessive growth of memory
computer structures), often produced in the phenotype of the
individual. During the last years, variants to GP like GE
appeared as a simpler optimization
process~\cite{ONeill2001}. In our approach, GE allows the generation
of a new set of optimized features by applying SR. This feature
generation is achieved thanks to the use of grammars that define the
rules for obtaining mathematical expressions. More concretely, we will
use grammars expressed in Backus Naur Form (BNF)~\cite{ONeill2001}.

A BNF specification is a set of derivation rules, expressed in the form:
\begin{verbatim}
<symbol>::=<expression>
\end{verbatim}

The rules are composed of sequences of terminals and
non-terminals. Symbols that appear at the left are non-terminals while
terminals never appear on a left side. In this case we can affirm that
\verb|<symbol>| is a non-terminal and, although this is not a complete
BNF specification, we can affirm also that \verb|<expression>| will be
also a non-terminal, since those are always enclosed between the pair
\verb|< >|. Therefore, in this case the non-terminal \verb|<symbol>|
will be replaced (indicated \verb|::=|) by an expression. The rest of
the grammar must define the different alternatives.

A grammar is represented by the 4-tuple ${N, T, P, S}$, being $N$ the
non-terminal set, $T$ is the terminal set, $P$ the production rules
for the assignment of elements on $N$ and $T$, and $S$ is a start
symbol that should appear in $N$. The options within a production rule
are separated by a ``$|$'' symbol.

\begin{figure}[ht]
        \centering
                \begin{verbatim}
N = {EXPR, OP, PREOP, VAR, NUM, DIG}
T = {+, -, *, /, sin, cos, exp, x, y, z, 
     0, 1, 2, 3, 4, 5, (, ), .}
S = {EXPR}
P = {I, II ,III ,IV ,V ,VI}
I   <EXPR> ::= <EXPR><OP><EXPR>
             | <PREOP>(<EXPR>)
             | <VAR>
II  <OP>   ::= + | - | * | /
III <PREOP>::= sin| cos | exp
IV  <VAR>  ::= x | y | z | <NUM>
V   <NUM>  ::= <DIG>.<DIG> | <DIG>
VI  <DIG>  ::= 0 | 1 | 2 | 3 | 4 | 5
                \end{verbatim}
        \caption{Example of a grammar in BNF format designed for
          symbolic regression}
        \label{fig:BnfExample}
\end{figure}

Figure~\ref{fig:BnfExample} represents an example of a grammar in BNF,
designed for symbolic regression. The final obtained expression will
consist of elements of the set of terminals $T$. These have been
combined with the rules of the grammar, as explained previously.
Also, grammars can be adapted to bias the search of the relevant
features because there is a finite number of options in each
production rule.

Regarding both the structure and the internal operators, GE works
exactly like a classic Genetic Algorithm (GA)~\cite{Back1997}. GE evolves a
population formed by a set of individuals, each one constituted by a
chromosome and a fitness value. In SR, the fitness value is usually a
regression metric like Root Mean Square Deviation (RMSD), Coefficient
of Variation (CV), Mean Squared Error (MSE), etc. In GE, a chromosome
is a string of integers. In the optimization process, GA operators
named selection, crossover and mutation are iteratively applied to
improve the fitness value of each individual. In order to compute the
fitness function for every iteration and to extract the mathematical
expression given by an individual, a decoding process is applied. We
refer the reader to~\cite{Goldberg1989} to understand the different GA
operators. In the following, we explain through an example the
decoding process performed in GE, since it clearly explains how better
features are automatically selected. Let us suppose that we have the
BNF grammar provided in Figure~\ref{fig:BnfExample} and the following
7-gene chromosome:

\begin{verbatim}
21-64-17-62-38-254-2
\end{verbatim}

According to Figure~\ref{fig:BnfExample}, the start symbol is
\verb|S={EXPR}|, hence the decoded expression will begin with this
non-terminal:
\begin{verbatim}
Solution = <EXPR>
\end{verbatim}
Now, we use the first gene of the chromosome (also called codon, equal
to 21 in the example) in rule \texttt{I} of the grammar. The number of
choices in that rule is 3. Hence, a mapping function (the modulus
operator) is applied:
\begin{verbatim}
21 MOD 3 = 0
\end{verbatim}
and the first option is selected \verb|<EXPR><OP><EXPR>|. The selected
option substitutes the decoded non-terminal. As a consequence, the
current expression is the following:
\begin{verbatim}
Solution = <EXPR><OP><EXPR>
\end{verbatim}
The process continues with the next codon, 64, which is used to decode
the first non-terminal of the current expression, namely,
\verb|<EXPR>|. Again, the mapping function is applied to rule
\texttt{I}:
\begin{verbatim}
64 MOD 3 = 1
\end{verbatim}
and the second option \verb|<PREOP>(<EXPR>)| is selected. So far, the
current expression is:
\begin{verbatim}
Solution = <PREOPR>(<EXPR>)<OP><EXPR>
\end{verbatim}
The next gene, 17, is now taken for decoding. At this point, the first
non-terminal in the current expression is \texttt{<PREOP>}. Therefore,
we apply the mapping function to rule \texttt{III}, which also has 3
different choices:
\begin{verbatim}
17 MOD 3 = 2
\end{verbatim}
and the third option \verb|exp| is selected. The resulting expression is
\begin{verbatim}
Solution = exp(<EXPR>)<OP><EXPR>
\end{verbatim}
Next codon, 62, decodes \verb|<EXPR>| with rule \texttt{I}:
\begin{verbatim}
62 MOD 3 = 2
\end{verbatim}
Value 2 means to select the third option, \verb|<VAR>|. The resulting
expression is:
\begin{verbatim}
Solution = exp(<VAR>)<OP><EXPR>
\end{verbatim}
Codon 38 decodes \verb|<VAR>| with rule \texttt{IV}:
\begin{verbatim}
38 MOD 4 = 2
\end{verbatim}
selecting the third option, non-terminal \texttt{z}:
\begin{verbatim}
Solution = exp(z)<OP><EXPR>
\end{verbatim}
Non-terminal \verb|<OP>| is decoded with codon 254 and rule \texttt{II}:
\begin{verbatim}
254 MOD 4 = 2
\end{verbatim}
This value selects the third option, terminal \verb|*|:
\begin{verbatim}
Solution = exp(z)*<EXPR>
\end{verbatim}
The last codon, decodes \verb|<EXPR>| with rule \texttt{I}:
\begin{verbatim}
2 MOD 3 = 2
\end{verbatim}
Value 2 selects the third option, non-terminal \verb|<VAR>|. So far, the current expression is:
\begin{verbatim}
Solution = exp(z)*<VAR>
\end{verbatim}
At this point, the decoding process has run out of codons. That is, we
have not arrived to an expression with terminals in all of its
components. In GE, the solution to this problem is to reuse codons
starting from the first one. In fact, it is possible to reuse the
codons more than once. This technique is known as wrapping and mimics
the gene-overlapping phenomenon in many organisms~\cite{Hemberg2013}.
Thus, applying wrapping to our example, the process goes back to the
first gene, 21, which is used to decode \verb|<VAR>| with rule
\texttt{IV}:
\begin{verbatim}
21 MOD 4 = 1
\end{verbatim}
This result selects the second option, non-terminal \verb|y|, giving
the final expression of the phenotype:
\begin{verbatim}
Solution = exp(z)*y
\end{verbatim}
As can be seen, the process does not only perform parameter
identification like in a classic regression method. In conjunction
with a well-defined fitness function, the evolutionary algorithm is
also computing an optimized set of features as mathematical
expressions by combining the set of parameters that best fits the
target system. Thus, GE is defining the optimal set of features that
will derive into the most accurate power model.

\subsection{Least absolute shrinkage and selection operator}

Tibshirani proposes the least absolute shrinkage and selection operator
algorithm (\emph{lasso})~\cite{LASSO1996} that minimizes residual summation
of squares according to the summation of the absolute value of the
coefficients that are less than constant.

The algorithm combines the favourable features of both subset
selection and ridge regression like stability, and offers a linear,
convex and derivable solution. \emph{Lasso} provides interpretable models
shrinking some of the coefficients and setting others to exactly zero
values for generalized regression problems.

For a given non-negative value of $\lambda$, the \emph{lasso} algorithm solves
the following problem:

\begin{eqnarray}
	\displaystyle\min_{\beta_{0}, \beta}{\left (\frac{1}{2N}\displaystyle\sum_{i=1}^N(y_i - \beta_{0} - x_{i}^T\beta)^2 + \lambda \displaystyle\sum_{j=1}^p |\beta_{j}|\right )} \label{eq:eqLASSO}
\end{eqnarray}

\noindent where:
\begin{itemize}

\item $\beta$: vector of $p$ components. \emph{Lasso} algorithm involves the
  $L^1$ norm of $\beta$

\item $\beta_{0}$: scalar value.

\item $N$: number of observations.

\item $y_i$: response at observation $i$.

\item $x_i$: vector of $p$ values at observation $i$.

\item $\lambda$: non-negative regularization parameter corresponding to
  one value of Lambda. The number of nonzero components of $\beta$
  decreases as $\lambda$ increases.

\end{itemize}

At the end, we combine the use of GE that generates the set of
relevant features with \emph{lasso} that computes the coefficients and the
independent term in the final linear model.

As a result, our GE+\emph{lasso} framework solves our optimization problem
that targets the generation of accurate power models for high-end
servers.

\section{Devised Methodology}
\label{probMod}

The fast and accurate modeling of complex systems is a relevant target
nowadays.  Modeling techniques allow designers to estimate the effects
of variations in the performance of a system.  Complex systems present
non-linear characteristics as well as a high number of potential
variables.  Also, the optimal set of features that impacts the system
performance is not well known as many mathematical relationships can
exist among them.

Hence, we propose a methodology that considers all these factors by
combining the benefits of both GE algorithms and
classical \emph{lasso} regressions.  This technique provides a generic and
effective modeling approach that could be applied to numerous problems
regarding complex systems, where the number of relevant variables or
their interdependence are not known.

\begin{figure*}
\centering
\includegraphics[width=0.70\textwidth]{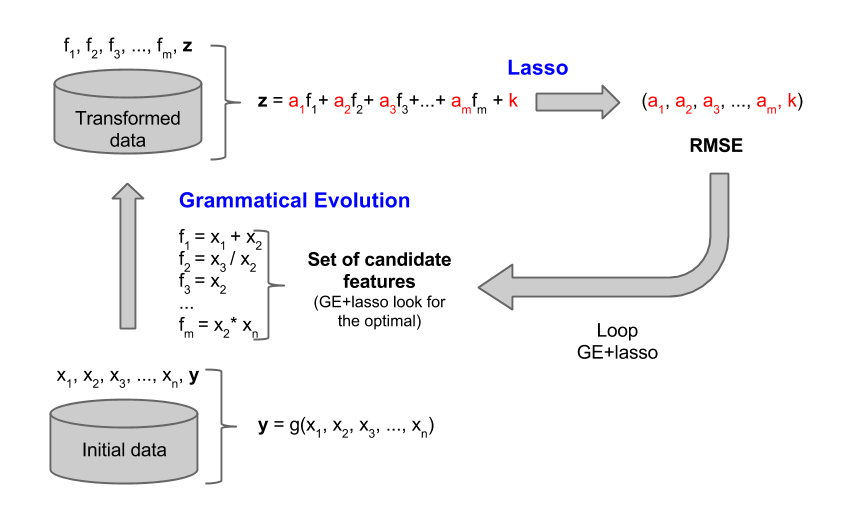}
\caption{Optimized modeling using GE+\emph{lasso} methodology.}
\label{fig:diagram}
\end{figure*}

Figure~\ref{fig:diagram} shows the proposed methodology approach for
the optimization of system modeling problem.  Detailed explanations of
the different phases are summarized in the following subsections.

\subsection{GE feature selection}

Given an extensive set of parameters that may cause an effect on
system performance, FE selects the optimal set that
best describes the system behavior.  Also, this technique, which is
provided by GE, avoids the inclusion of irrelevant
features while incorporating correlations and combinations of
representative variables.

The input to our approach consists of a vector of initial data that
includes the entire set of variables $x_n$ extracted from the system.
\begin{eqnarray}
\vec{y} = g_1 (x_1, x_2, x_3, \dots, x_n)
\end{eqnarray}
All these parameters are entered in the GE algorithm to
start the optimization process.\\

Each individual of the GE encodes its own set of candidate
features $f_1, f_2, f_3, \dots, f_m$.  The candidate features follow the rules
imposed by a BNF grammar allowing the occurrence of a wide variety of
operations and operands to favor building optimal sets of features.
Figure~\ref{fig:BnfGenericmodeling} shows an example of a BNF grammar for this
approach.

\begin{figure}[ht]
        \centering
                \begin{verbatim}
<list_features> ::= <feature>
                  | <feature>;<list_features>
<feature>       ::= (<feature><op><feature>)
                  | <preop>(<feature>)
                  | <var>
<op>            ::= + | - | * | / 
<preop>         ::= exp | sin | cos | ln
<var>           ::= x[0] | x[1] | x[2] | x[3] 
                  | x[4] | x[5] | ... | x[n]
                \end{verbatim}
                \caption{Grammar in BNF format. $x$ variables, with $i=0
                  \ldots n$, represent each parameter obtained from the system.}
        \label{fig:BnfGenericmodeling}
\end{figure}

This grammar provides the operations $+$, $-$, $*$, $/$ and
preoperators $exp$, $sin$, $cos$, $ln$.  The space of solutions is
easily modified by incorporating a broader set of relationships
between operands to the BNF grammar.

The output of the GE stage consists of a matrix
that includes all the candidate features provided by individuals.
Each individual output vector has its own set of $m$ candidate
features that intends to minimize the fitness function provided for
the system optimization.
\begin{eqnarray}
\vec{z} = g_2 (f_1, f_2, f_3, \dots, f_m)
\end{eqnarray}

\subsection{\emph{Lasso} generic model generation}

Modeling procedures usually intend to interpret systems' behavior.
They have the purpose of acquiring additional knowledge from the final
models once these have been derived.  Linearity, convexity and
differentiability offered by \emph{lasso} classic regression helps modeling
to be a more explanatory and repeatable process.  In addition, whereas
GE is able to find complex symbolic expressions, GE
does not perform well in parameter identification, mainly because the
exploration of real numbers is not easily representable in BNFs.  Due
to these facts, we have included \emph{lasso} algorithm in our methodology in
order to manage the coefficient generation of the system model.

As can be seen in Figure~\ref{fig:diagram}, each individual of the
GE provides a set of candidate features to
\emph{lasso}. This classical regression is in charge of deriving the
optimized model for each individual by solving the following equation.
\begin{eqnarray}
\vec{z} = a_1 f_1 + a_2 f_2 + a_3 f_3 + \dots + a_m f_m + k
\end{eqnarray}

\emph{Lasso} offers the set of optimized coefficients ($a_1$, $a_2$, $a_3$,
$\dots$, $a_m$, $k$) for each individual that minimizes the fitness
function.  This process provides the goodness of each individual.  All
this information feeds back the GE algorithm to
generate the next population of individuals through selection,
crossover and mutation, creating a loop.  The loop continues executing
until it completes the number of generations defined by the
GE. This process results in the set of models that
best fits system performance.

\subsection{Fitness evaluation}
As our main target is to build accurate models, our fitness function
includes the error resulting in the estimation process. The fitness
function presented in~\ref{eq:fitness} leads the evolution to obtain
optimized solutions thus minimizing Root Mean Square Error or RMSD.
\begin{eqnarray}
\label{eq:fitness}
F & = & \sqrt{\frac{1}{N} \cdot \sum_{n} {e_\mathrm{n}}^2} \\
e_\mathrm{n} & = & | P(n) - \widehat{P}(n) |, \qquad 1 \leq n \leq N \qquad
\end{eqnarray}

Estimation error $e_\mathrm{n}$ represents the deviation between the
measure given by system monitoring $P$, and the estimation obtained by
the model $\widehat{P}$. $n$ represents each sample of the entire set
of $N$ samples used to train the algorithms.

\section{Case Study}
\label{caseStudy}

In this section we describe a particular case study for the
application of the devised methodology presented in
Section~\ref{probMod}.  The problem to be solved is the fast and
accurate estimation of the power consumption in virtualized enterprise
servers performing Cloud applications.  Our power model considers
heterogeneity of servers, as well as specific technological features
and non-traditional parameters of the target architecture that affect
power consumption. Hence, we propose our modeling technique that
considers all these factors by combining the benefits of both
GE algorithms and classical \emph{lasso} regressions.

Firstly, a GE algorithm is applied to extract those
relevant features that best describe power consumption
sources. Features may also include correlations and combinations of
representative variables due to FE performed by
GE.  Then, the \emph{lasso} algorithm takes the optimal
set of features in order to infer an expression that characterizes the
power behavior of the target architecture of a Cloud server. As a
result, we derive a highly accurate, linear and convex power model,
targeting a specific server architecture, that is automatically
generated by our evolutionary methodology.

We apply our methodology described in Section~\ref{probMod} to real
measures gathered from a high-end Cloud server in order to infer an
accurate power model. Also, we provide an experimental scenario for
various workloads with the purpose of building and testing our
methodology.

\subsection{Data compilation}
Data have been collected gathering real measures from a Fujitsu RX300
S6 server based on an Intel Xeon E5620 processor. This high-end server
has a RAM memory of 16GB and is running a 64bit CentOS 6.4 OS
virtualized by the QEMU-KVM hypervisor.  Physical resources are
assigned to four KVM virtual machines, each one with a CPU core and a
4GB RAM block.

The power consumption of a high-end server usually depends on several
factors that affect both dynamic and static
behavior~\cite{parrobaSEB}. Our proposed case study takes into account
the following 7 variables:

\begin{itemize}
\item \emph{Ucpu}: CPU utilization  (\%)
\item \emph{Tcpu}: CPU temperature  (Kelvin)
\item \emph{Fcpu}: CPU frequency (GHz)
\item \emph{Vcpu}: CPU voltage (V)

\item \emph{Umem}: Main memory utilization (Memory accesses per cycle)
\item \emph{Tmem}: Main memory temperature (Kelvin)

\item \emph{Fan}: Fan speed (RPM)
\end{itemize}

Power consumption is measured with a current clamp with the aim of
validating our approach.  CPU and main memory utilization are
monitored using hardware counters collected with \emph{perf}
monitoring tool.  On board sensors are checked via IPMI to get both
CPU and memory temperatures and fan speed.  CPU frequency and voltage
are monitored via \emph{cpufreq-utils} Linux package. To build a model
that includes power dependance with these variables, we use this
software tool to modify CPU DVFS modes during workload execution.
Also room temperature has been modified in run-time with the goal of
finding non-traditional consumption sources that are influenced by
this variable.

\subsection{Experimental workload}
We define three workload profiles (i) synthetic, (ii) Cloud and (iii)
HPC over Cloud as they emulate different utilization patterns that
could be found in typical Cloud infrastructures.

\subsubsection{Synthetic benchmarks}
The use of synthetic load allows to specifically stress different
server resources.  The importance of using synthetic load is to
include situations that do not meet the actual real workloads that we
have selected. Thus, the range of possible values of the different
variables is extended in order to adapt the model to fit future
workload characteristics and profiles.
\emph{Lookbusy}\footnote{http://www.devin.com/lookbusy/} stresses
different CPU hardware threads to a certain utilization avoiding
memory or disk usage.  The memory subsystem is also stressed
separately using a modified version of
\emph{RandMem}\footnote{http://www.roylongbottom.org.uk}. We have
developed a program based on this benchmark to access random memory
regions individually, with the aim of exploring memory
performance. \emph{Lookbusy} and \emph{RandMem} have been executed, in
a separated and combined fashion, onto 4 parallel Virtual Machines
that consume entirely the available computing resources of the server.

On the other hand, real workload of a Cloud data center is represented
by the execution of \emph{Web Search}, from
\emph{CloudSuite}\footnote{http://parsa.epfl.ch/cloudsuite}, as well
as by \emph{SPEC\_CPU2006 mcf} and \emph{SPEC\_CPU2006
  perlbench}~\cite{Henning:2006:SCB:1186736.1186737}.

\subsubsection{Cloud workload}
\emph{Web Search} characterizes web search engines, which are typical
Cloud applications. This benchmark processes client requests by
indexing data collected from online sources. Our \emph{Web Search}
benchmark is composed of three VMs performing as index serving nodes
(ISNs) of Nutch 1.2. Data are collected in the distributed file system
with a data segment of 6 MB, and an index of 2 MB that is crawled from
the public Internet. One of this ISNs also executes a Tomcat 7.0.23
frontend in charge of sending index search requests to all the
ISNs. The frontend also collects ISNs responses and sends them back to
the requesting client. Client behavior is generated by Faban $0.7$
performing in a fourth VM.  Resource utilization depends
proportionally on the number of clients accessing \emph{Web
  Search}. Our number of clients configuration ranges from 100 to 300
to expose more information about the application performance.  The
four VMs use all the memory and CPU resources available in each
server.

\subsubsection{HPC over Cloud}
In order to represent HPC over a Cloud computing infrastructure, we
choose \emph{SPEC\_CPU2006 mcf} and \emph{perlbench} as they are
memory and CPU-intensive, and CPU-intensive applications,
respectively.  \emph{SPEC\_CPU2006 mcf} consists on a network simplex
algorithm accelerated with a column generation that solves large-scale
minimum-cost flow problems. On the other hand, a mail-based benchmark
is performed by \emph{SPEC\_CPU2006 perlbench}. This program applies a
spam checking software to randomly generated email messages. Both SPEC
applications are run in parallel in 4 VMs using entirely the available
resources of the server.

Instead of restricting the use of synthetic workloads only for
training the algorithms, and limiting the use of real Cloud benchmarks
exclusively for testing, we have used both workloads for the two
purposes.  This procedure provides automation for the progressive
incorporation of additional benchmarks to the model.

For each run of the combined GE+\emph{lasso} approach, we randomly select
50\% of each data set (synthetic, \emph{Web Search},
\emph{SPEC\_CPU2006 mcf} and \emph{perlbench}) for training and the
remaining 50\% for testing stage.  This technique validates the
variability and versatility of the system, by analyzing the occurrence
of local minima in optimization scenarios.

\section{Experimental results}
\label{results}

As we stated in section~\ref{caseStudy}, tests have been conducted
gathering real data from our Fujitsu server.  Our experiments present
high variability for the different features compiled from the server.

\begin{itemize}

\item CPU operation frequency (\emph{Fcpu}) is fixed to
  $f_{1}=1.73$~GHz, $f_{2}=1.86$~GHz, $f_{3}=2.13$~GHz,
  $f_{4}=2.26$~GHz, $f_{5}=2.39$~GHz and $f_{6}=2.40$~GHz; thus
  modifying CPU voltage (\emph{Vcpu}) from 1.73~V to 2.4~V.

\item Room temperature has been modified in run-time, from
  17$^{\circ}$C to 27$^{\circ}$C. Therefore, temperatures of CPU and
  memory (\emph{Tcpu} and \emph{Tmem}) range from 306~K to 337~K, and
  from 298~K to 318~K respectively.

\item CPU and memory utilizations (\emph{Ucpu} and \emph{Umem}) take
  values from 0\% to 100\% and from 0 to 0.508 memory accesses
  (cache-misses) per CPU cycle respectively.

\item Finally, due to both room temperature, and CPU and memory
  utilization variations, fan speed values (\emph{Fan}) range from
  3540~RPM to 7200~RPM.
\end{itemize}

Data have been split into training and testing sets. Training stage
performs feature selection and builds the power model according to our
grammar and fitness function.  Next, the testing stage examines the
power model accuracy.  The algorithm proposed by our methodology is
executed completely 20 rounds using the same grammar and fitness
function configuration.  For each run, we randomly select 50\% of each
data set for training and 50\% for testing stage, thus obtaining 20
final power models. This procedure validates the variability and
versatility of the system, by analyzing the occurrence of local minima
in optimization scenarios.

\subsection{Algorithm setup}
\label{sec:setup}

\subsubsection{GE setup parameters}
We use GE to obtain a set of candidate features that best describe our
optimization problem. To obtain adequate solutions we tune the
algorithm using the following parameters:

\begin{itemize}
\item Population size: 250 individuals
\item Number of generations: 3000
\item Chromosome length: 100 codons
\item Mutation probability: inversely proportional to the number of rules, $1/4$  in our case.
\item Crossover probability: 0.9
\item Maximum wraps: 3
\item Codon size: 256
\item Tournament size: 2 (binary)
\end{itemize}

As we strictly seek for simple combination of features, our proposed
BNF grammar only provides the operations $+ | - | * | /$.  The space
of solutions is easily increased by incorporating more complex
relationships between operands to the BNF grammar.
Figure~\ref{fig:BnfcaseStudy} shows the BNF grammar proposed for this
case study.

\begin{figure}[ht]
        \centering
                \begin{verbatim}
<list_features> ::= <feature>
                  | <feature>;<list_features>
<feature>       ::= (<feature><op><feature>)
                  | <var>
<op>            ::= + | - | * | /
<var>           ::= x[0] | x[1] | x[2] | x[3] 
                  | x[4] | x[5] | x[6]
                \end{verbatim}
                \caption{Grammar in BNF format. $x$ variables, with $i=0
                  \ldots 6$, represent processor and memory temperatures, fan
                  speed, processor and memory utilization percentages, processor
                  frequency and voltage, respectively.}
        \label{fig:BnfcaseStudy}
\end{figure}

\subsubsection{Lasso setup parameters}
We use the \emph{lasso} algorithm to obtain a set of candidate solutions with
low error, when compared with the real power consumption measures in
order to solve our optimization modeling problem. \emph{Lasso} setup
parameters are the following:

\begin{itemize}
\item Number of observations: 100
\item $\lambda$ regularization parameter: Geometric sequence of 100 values, the largest just sufficient to produce zero coefficients.
\item $\lambda$ regularization parameter: $1 \cdot 10^{-4}$
\end{itemize}

\subsection{Training stage}
We have performed variable standardization for every feature (in the range $[1,
2]$) to assure the same probability of appearance for all the variables and to
enhance the GE symbolic regression. Experiments with more than 4 features do
not provide better values for RMSD. Hence, we have bounded their occurrence to
a maximum of 4 by penalizing higher number of features in our fitness
evaluation function. This also facilitates the generation of simpler models,
which are faster and easy to be applied, in order to be used for real-time
power optimizations.  Table~\ref{tab:phenotype} shows phenotypes of each
feature combined with the coefficients provided by \emph{lasso} that are obtained for
20 complete executions of our methodology algorithm.  Fitness results, that
correspond to the RMSD between measured and estimated power consumption (see
Equation~\ref{eq:fitness}), are shown in Table~\ref{tab:RMSD} for the training
stage.  Both Table~\ref{tab:phenotype} and Table~\ref{tab:RMSD} present the
results for the best model of each execution.

\begin{table*} [!ht]
  \scriptsize
  \centering
  \caption {Power models obtained by combining GE features and \emph{lasso}
    coefficients for 20 executions}
  \begin{tabular}{ll}
  \hline
  Run & $a_1$ $\cdot$ $f_1$ + $a_2$ $\cdot$ $f_2$ + $a_3$ $\cdot$ $f_3$ + $a_4$ $\cdot$ $f_4$ + K \\
  \hline
1 & 0.288 $\cdot$ Tcpu \\  & + 0.127 $\cdot$ (((Tcpu*Ucpu)-Umem)*Fan) \\  & + 0.220 $\cdot$ (Fan*Tmem) \\  & + -0.450 $\cdot$ Fan + 1.043 \\ \hline
2 & 0.173 $\cdot$ Ucpu \\  & + 0.438 $\cdot$ Tcpu \\  & + -0.209 $\cdot$ Fan \\  & + 0.070 $\cdot$ (Tmem/(Umem/(Fan*Tmem))) + 0.636 \\ \hline
3 & 0.256 $\cdot$ (Fan/(Ucpu/Tmem)) \\  & + 0.346 $\cdot$ Ucpu \\  & + -0.694 $\cdot$ (Fan/Tcpu) + 1.151 \\ \hline
4 & -0.376 $\cdot$ Tmem \\  & + -0.033 $\cdot$ ((((Fan/Tcpu)/(Fcpu+Fcpu))/((Fan/(Vcpu+Umem))/Ucpu))*Fan) \\  & + 0.606 $\cdot$ ((Fan/((Umem+(Fan+(Fcpu/Fcpu)))*(Fan/Ucpu)))+(Fan+Tmem)) \\  & + 0.786 $\cdot$ ((Fcpu-(Fcpu+Fan))/Tcpu) + 0.810 \\ \hline
5 & 0.181 $\cdot$ Ucpu \\  & + 0.254 $\cdot$ (Fan*Tmem) \\  & + 0.378 $\cdot$ Umem \\  & + -0.345 $\cdot$ (((Umem+Umem)*Fan)/Tcpu) + 0.939 \\ \hline
6 & 0.483 $\cdot$ (Ucpu-Fan) \\  & + 0.030 $\cdot$ ((Tmem+Fan)*((Fan-(Tmem/((Ucpu+Vcpu)+(Fan+Fan))))*(Fan*Fan))) \\  & + 0.220 $\cdot$ Tmem \\  & + 0.430 $\cdot$ (Tcpu/Ucpu) + 0.402 \\ \hline
7 & 0.506 $\cdot$ Tcpu \\  & + 0.195 $\cdot$ ((Ucpu/(Vcpu+(Tmem-Umem)))*Vcpu) \\  & + -0.319 $\cdot$ Fan \\  & + -0.199 $\cdot$ (((Fan+Umem)*((Umem-Tmem)/Tcpu))*Fan) + 0.704 \\ \hline
8 & 0.084 $\cdot$ (Ucpu/Vcpu) \\  & + 0.473 $\cdot$ Tmem \\  & + 0.499 $\cdot$ (Ucpu/(Ucpu*(((Fcpu-Vcpu)+Tmem)/Tcpu))) \\  & + -0.019 $\cdot$ (Fan-(((Fan+Vcpu)*Tcpu)*((Ucpu*Tmem)-(Vcpu-Fcpu)))) + 0.046 \\ \hline
9 & 0.927 $\cdot$ Ucpu \\  & + -0.380 $\cdot$ Fan \\  & + 0.232 $\cdot$ (((Tmem*((Fan+Umem)-Ucpu))+(Tcpu-(Ucpu*Umem)))-Ucpu) \\  & + 0.180 $\cdot$ Tcpu + 0.365 \\ \hline
10 & -0.073 $\cdot$ Tmem \\  & + 0.106 $\cdot$ (((Tmem+Fan)*Fan)-Umem) \\  & + 0.194 $\cdot$ (Ucpu+Tmem) \\  & + 0.437 $\cdot$ (Tcpu-Fan) + 0.665 \\ \hline
11 & -0.117 $\cdot$ (Tmem*(Ucpu-(Tmem*Fan))) \\  & + 0.317 $\cdot$ Ucpu \\  & + 0.377 $\cdot$ (Tcpu-Fan) + 0.810 \\ \hline
12 & -0.070 $\cdot$ Umem \\  & + 0.174 $\cdot$ Ucpu \\  & + 0.647 $\cdot$ (Tcpu/Tmem) \\  & + 0.647 $\cdot$ Tmem + -0.318 \\ \hline
13 & 0.291 $\cdot$ (Tmem+Fan) \\  & + -0.409 $\cdot$ (Fan/Tcpu) \\  & + 0.234 $\cdot$ Tcpu \\  & + 0.423 $\cdot$ (Ucpu/(Tmem+Umem)) + 0.442 \\ \hline
14 & 0.093 $\cdot$ (Ucpu+(Ucpu+(Tmem*Tmem))) \\  & + -0.019 $\cdot$ ((Tcpu-((Tmem*Fan)-Vcpu))-Vcpu) \\  & + -0.081 $\cdot$ (Tmem+Umem) \\  & + 0.462 $\cdot$ Tcpu + 0.526 \\ \hline
15 & -0.004 $\cdot$ Fcpu \\  & + 0.380 $\cdot$ (Ucpu/(Umem+Fan)) \\  & + 0.054 $\cdot$ (Tmem*(Tmem+Fan)) \\  & + 0.454 $\cdot$ Tcpu + 0.347 \\ \hline
16 & -0.010 $\cdot$ Fan \\  & + -0.155 $\cdot$ (((Fan/Tmem)-(Tmem/Ucpu))*Fan) \\  & + 0.282 $\cdot$ Ucpu \\  & + 0.417 $\cdot$ Tcpu + 0.393 \\ \hline
17 & 0.242 $\cdot$ (Fan*(Tmem/Ucpu)) \\  & + 0.396 $\cdot$ (Tcpu-Fan) \\  & + 0.001 $\cdot$ Fcpu \\  & + 0.344 $\cdot$ Ucpu + 0.508 \\ \hline
18 & 0.448 $\cdot$ Tmem \\  & + -0.178 $\cdot$ Umem \\  & + -0.221 $\cdot$ (((((Tcpu/(Vcpu/(Fcpu-Vcpu)))-Ucpu)+Fan)/Tmem)-(Tcpu-(Tmem-(Tcpu+Fan)))) \\  & + 0.100 $\cdot$ (Umem/Fan) + 0.271 \\ \hline
19 & 0.134 $\cdot$ Ucpu \\  & + 0.241 $\cdot$ (Tmem*Fan) \\  & + 0.066 $\cdot$ Ucpu \\  & + -0.403 $\cdot$ ((Fan-Tcpu)/Umem) + 0.653 \\ \hline
20 & -0.433 $\cdot$ (((Fan-(Ucpu+Umem))/Fan)-(Tcpu+Fan)) \\  & + -0.295 $\cdot$ Umem \\  & + -0.102 $\cdot$ Fan \\  & + 0.235 $\cdot$ (((Tmem-Umem)-Ucpu)+Fan) + 0.184 \\
\hline
  \end{tabular}
  \label{tab:phenotype}
\end{table*}

\begin{table*} [t]
  \centering
  \caption {RMSD and Average testing error percentages for 20 executions}
  \begin{tabular}{cccccccc}
  \hline
  Run & Train (RMSD) & Validation (RMSD) & Synthetic (\%) & mcf (\%) & perlb (\%)  & WebSearch (\%)  & Total (\%)  \\
  \hline
  1 & 0.1069 & 0.1068 & 3.985 & 4.097 & 4.463 & 4.147 & 4.173 \\
  2 & 0.1068 & 0.1067 & 3.984 & 4.099 & 4.463 & 4.110 & 4.164 \\
  3 & 0.1070 & 0.1068 & 3.995 & 4.110 & 4.504 & 4.145 & 4.189 \\
  4 & 0.1070 & 0.1071 & 4.007 & 4.085 & 4.469 & 4.155 & 4.179 \\
  5 & 0.1069 & 0.1069 & 3.991 & 4.106 & 4.494 & 4.113 & 4.176 \\
  6 & 0.1071 & 0.1068 & 3.988 & 4.085 & 4.459 & 4.153 & 4.171 \\
  7 & 0.1070 & 0.1072 & 3.995 & 4.042 & 4.462 & 4.101 & 4.150 \\
  8 & 0.1071 & 0.1072 & 3.994 & 3.996 & 4.559 & 4.101 & 4.162 \\
  9 & 0.1072 & 0.1072 & 4.033 & 3.884 & 3.990 & 4.059 & 3.991 \\
  10 & 0.1067 & 0.1072 & 4.052 & 3.894 & 3.969 & 4.031 & 3.986 \\
  11 & 0.1073 & 0.1075 & 4.023 & 3.926 & 3.963 & 4.063 & 3.994 \\
  12 & 0.1071 & 0.1076 & 4.098 & 3.896 & 3.951 & 4.030 & 3.994 \\
  13 & 0.1070 & 0.1070 & 4.073 & 3.939 & 4.173 & 4.243 & 4.107 \\
  14 & 0.1072 & 0.1072 & 4.088 & 3.935 & 4.174 & 4.184 & 4.096 \\
  15 & 0.1071 & 0.1070 & 4.083 & 3.922 & 4.161 & 4.246 & 4.103 \\
  16 & 0.1071 & 0.1070 & 4.060 & 3.937 & 4.164 & 4.217 & 4.095 \\
  17 & 0.1079 & 0.1057 & 3.951 & 4.136 & 4.208 & 4.056 & 4.088 \\
  18 & 0.1081 & 0.1060 & 3.981 & 4.171 & 4.180 & 4.050 & 4.095 \\
  19 & 0.1082 & 0.1060 & 3.953 & 4.190 & 4.224 & 4.212 & 4.145 \\
  20 & 0.1082 & 0.1059 & 3.974 & 4.205 & 4.178 & 4.074 & 4.108 \\
  \hline
  \end{tabular}
  \label{tab:RMSD}
\end{table*}

As can be seen in Table~\ref{tab:phenotype}, power model solutions
combine features that correspond to a single variable with others that
merge a combination of several parameters.  On the one hand, there are
single variable features that appear in up to 50\% of the power model
solutions.  This shows that there are linear dependencies with certain
parameters, as \emph{Ucpu}, \emph{Tpcu}, and \emph{Tmem} that are consistent regardless of
the workload that is used for training and testing.  On the other
hand, variables as \emph{Vcpu}, \emph{Fcpu} and \emph{Umem} are seldom treated as a feature
in the model solutions.  However, they systematically appear when
combined with other variables.  These results show how there exist
input parameters that are not relevant for the modeling or they are
correlated to other features, and their inclusion could decrease the
model accuracy.  Model training for run 10 shows the lowest RMSD error
of 0.1067.

\subsection{Model testing}
At this stage, we analyze the quality of the models that we have
simultaneously tested for the 20 complete executions of our
methodology algorithm. Results are also analyzed particularly for the
testing data that corresponds to each benchmark dataset in order to
verify the estimation reliability of the models for different
workloads.  Table~\ref{tab:RMSD} shows testing average error
percentages particularized for the different benchmark data sets.
These values have been obtained according to the following
formulation:
\begin{eqnarray}
\label{eq:avgError}
e_\mathrm{AVG} = \sqrt{\frac{1}{N} \cdot \sum_{n} {(\frac{| P(n) - \widehat{P}(n) | \cdot 100} {P(n)})}^2}, 1 \leq n \leq N
\end{eqnarray}

\noindent where $P$ is the power measure given by the current clamp
and $\widehat{P}$ is the power estimated by the model phenotype. $n$
represents each sample of the entire set of $N$ samples.

Total average error for the testing dataset shows lowest error of
3.98\% (as shown in Table~\ref{tab:RMSD}).  Best testing error
corresponds to the solution with lower training error.  Solutions can
be broken down for those samples that belong to different tests,
achieving testing errors of 4.052\%, 3.894\%, 3.969\% and 4.031\% for
synthetic, \emph{SPEC\_CPU2006 mcf}, \emph{SPEC\_CPU2006 perlbench} and \emph{WebSearch} workloads respectively.  This
fact confirms that our methodology works well for our scenario,
extracting optimized sets of features and coefficients that are
consistent even for 20 runs with random selection of both training and
testing dataset.
 
Our methodology application shows very accurate testing results for
all of the whole executions ranging from 3.98\% to 4.18\%.  The
obtained results are robust, as they have been obtained for a
heterogeneous mix of workloads so the power models are not
workload-dependant.  According to these results, we can infer that our
methodology is effective for performing feature selection and building
accurate multi-parametric, linear, convex and differentiable power
models for high-end Cloud servers.  This technique can be considered
as a starting point for implementing energy optimization policies for
Cloud computing facilities.

\section{Conclusions}
\label{conclusions}

This paper has presented a novel work in the field of FE
and SR for the automatic inference of
accurate models. Resulting models include combination and correlation
of variables due to the FE and SR
performed by GE.  Therefore, the models incorporate
the optimal selection of representative features that best describe
the target problem while providing linearity, convexity and differentiability
characteristics.

As a proof of concept, the devised methodology has been applied to a
current computing problem, the power modeling of high-end servers in a
Cloud environment. In this context, the proposed methodology has shown
relevant benefits with respect to state-of-the-art approaches, like
better accuracy and the opportunity to consider a broader number of
input parameters that can be exploited by further power optimization
techniques.

\section{Acknowledgments}
\label{acknowledgment}
Research by Marina Zapater has been partly supported by a PICATA predoctoral fellowship of the Moncloa Campus of International Excellence (UCM-UPM).  This work has been partially supported by the Spanish Ministry of Economy and Competitiveness, under contracts TIN2008-00508, TEC2012-33892 and IPT-2012-1041-430000, and INCOTEC. The authors thankfully acknowledge the computer resources, technical expertise and assistance provided by the Centro de Supercomputación y Visualización de Madrid (CeSViMa).

\bibliographystyle{spmpsci}      
\bibliography{bib/thermal}   


\end{document}